%% file: OTSMain_pre_print.tex
\documentclass[preprint,journal]{IEEEtran}

\usepackage{amsthm}
    \usepackage{graphicx}
    \usepackage{epstopdf}
    \usepackage[subrefformat=parens, caption=false, font=footnotesize]{subfig}
    \usepackage{booktabs}
    \usepackage{array}

    \usepackage[acronym,toc]{glossaries}
    \usepackage[nolist,nohyperlinks]{acronym}
    \loadglsentries[\acronymtype]{setup/acro}

    \usepackage{amsmath,amssymb,amsfonts}
    \usepackage{algorithmic}
    
    \usepackage{textcomp}
    \usepackage[svgnames]{xcolor}
    \usepackage{blindtext}
    \usepackage{outlines}
    \usepackage[normalem]{ulem}
    \usepackage{algorithm2e}

    \usepackage{cite}
    
    \usepackage{comment}
    \usepackage[section]{placeins}

\begin{document}

\input{setup/header.tex}

\input{sections/0_abstract.tex}
\acresetall

\input{sections/1_intro.tex}      
\input{sections/2_sec.tex}        
\input{sections/3_sec.tex}         
\input{sections/4_sec.tex}            
\input{sections/5_sec.tex}          
\input{sections/6_conc.tex}

\bibliographystyle{IEEEtran}
\footnotesize\bibliography{bib/refs.bib}

\end{document}

%% file: setup/acro.tex
\begin{acronym}

    \acro{AHT}[AHT]{active hypothesis testing}
    \acro{BER}[BER]{bit error rate}
    \acro{CHP2}[CHP2]{Communications and High-Precision Positioning}
    \acro{CPM}[CPM]{continuous phase modulation}
    \acro{DAS}[DAS]{Deterministic Adaptive Strategy}
    \acro{FFRED}[FFRED]{far-field radiated emission design}
    \acro{GDoP}[GDoP]{geometric dilution of precision}
    \acro{GPS}[GPS]{global positioning system}
    \acro{INR}[INR]{interference to noise ratio}
    \acro{ITS}[ITS]{intelligent transport systems}
    \acro{JRCPNT}[JRCPNT]{joint radar, communications, positioning, navigation, and timing}
    \acro{MCMUR}[MCMUR]{multiple-channel, multiple-user receiver}
    \acro{MISO}[MISO]{multiple-input single-output}
    \acro{MSE}[MSE]{mean squared error}
    \acro{NTP}[NTP]{network timing protocol}
    \acro{OFDM}[OFDM]{orthogonal frequency-division multiplexing}
    \acro{ORS}[ORS]{Open-Loop Randomized Strategy}
    \acro{PARC}[PARC]{phase-attached radar communications}
    \acro{PCFM}[PCFM]{polyphase-coded frequency modulated}
    \acro{PNT}[PNT]{positioning, navigation, and timing}
    \acro{SAR}[SAR]{synthetic aperture radar}
    \acro{SDRs}[SDRs]{software-defined radios}
    \acro{SIMO}[SIMO]{single-input multiple-output}
    \acro{SISO}[SISO]{single-input single-output}
    \acro{SNR}[SNR]{signal-to-noise ratio}
    \acro{SINR}[SINR]{signal-to-interference-plus-noise ratio}
    \acro{INR}[INR]{interference-to-noise ratio}
    \acro{STAP}[STAP]{space-time adaptive processing}
    \acro{SWaP-C}[SWaP-C]{size, weight, power, and cost}
    \acro{THoRaCs}[THoRaCs]{tandem-hopped radar communications}
    \acro{AoA}[AoA]{angle-of-arrival}
    \acro{ToA}[ToA]{time-of-arrival}
    \acro{ToF}[ToF]{time-of-flight}
    \acrodefplural{ToF}[ToFs]{times-of-flight}
    \acro{TDoA}[TDoA]{time-difference-of-arrival}
    \acrodefplural{TDoA}[TDoAs]{time-differences-of-arrival}
    \acro{TWR}[TWR]{two-way ranging}
    \acro{UAM}[UAM]{urban air mobility}
    \acro{V2V}[V2V]{vehicle-to-vehicle}
    \acro{V2I}[V2I]{vehicle-to-infrastructure}
    \acro{CDMA}[CDMA]{code-division multiple access}
    \acro{TDMA}[TDMA]{time-division multiple access}
    \acro{QPSK}[QPSK]{quadrature phase shift keying}
    \acro{QAM}[QAM]{quadrature amplitude modulation}
    \acro{CFO}[CFO]{carrier frequency offset}
    \acro{DoF}[DoFs]{degrees-of-freedom}
    \acro{MMSE}[MMSE]{minimum mean square error}
    \acro{OTS}[OTS]{optical time synchronization}
    \acro{TDEV}[TDEV]{time deviation}
    
\end{acronym}

%% file: setup/header.tex

    \title{Distributed Coherent Beamforming at 60 GHz Enabled by Optically-Established Coherence}

    \author{\IEEEauthorblockN{Drake Silbernagel$^1$, Yu Rong$^1$, Isabella Lenz$^1$,
        Prithvi Hemanth$^1$, 
        Carl Morgenstern$^1$,
        Owen Ma$^1$,
        Nolan Matthews$^2$, 
        Nader Zaki$^2$,
        Kyle W.~Martin$^3$,
        John D.~Elgin$^4$, 
        Jacob Holtom$^5$,
        Daniel W.~Bliss$^1$,
        Kimberly Frey$^4$ 
        }}

    \IEEEoverridecommandlockouts
  
    \maketitle
    \IEEEpubidadjcol
    \long\def\symbolfootnote[#1]#2{\begingroup\def\thefootnote{\fnsymbol{footnote}}
    %
    \footnote[#1]{#2}\endgroup}
    \symbolfootnote[0]{\hrule\vspace{1.5mm} 
    \noindent\begin{tabular}{ p{0.01cm} p{8cm} }
        $^1$ & Center for Wireless Information Systems and Computational Architectures (WISCA), Arizona State University, Tempe, AZ, 85281, USA \vspace{-2.1mm} \\ 
        $^2$ & Space Dynamics Laboratory, North Logan, UT  84341, USA \\
        $^3$ & Blue Halo, Albuquerque, NM  87123, USA \\
        $^4$ & Space Vehicles Directorate, Air Force Research Laboratory, Kirtland Air Force Base, Albuquerque, NM 87117, USA \\
        $^5$ & DASH Tech Integrated Circuits, Phoenix, AZ, USA
    \end{tabular}

    \vspace{2mm}
    This material is based on research sponsored by AFRL under agreement number FA9453-20-2-0001. 
    Approved for public release, distribution is unlimited. Public Affairs release approval \#AFRL20253316. The views expressed are those of the authors and do not reflect the official policy or position of the Department of the Air Force, the Department of Defense, or the U.S. government. }

%% file: sections/0_abstract.tex
\begin{abstract}We implement and experimentally demonstrate a 60 GHz distributed system leveraging an optical time synchronization system that provides precise time and frequency alignment between independent elements of the distributed mesh. Utilizing such accurate coherence, we perform receive beamforming with interference rejection and transmit nulling. In these configurations, the system achieves a coherent gain over an incoherent network of $N$ nodes, significantly improving the relevant signal power ratios. Our system demonstrates extended array phase coherence times, enabling advanced techniques. Results from over-the-air experiments demonstrate a 14.3 dB signal-to-interference-plus-noise improvement in interference-laden scenarios with a contributing 13.5 dB null towards interference in receive beamforming. In transmit nulling, a signal-to-noise ratio (SNR) gain of 7.9 dB is measured towards an intended receiver while maintaining an SNR reduction of 8.9 dB at another receiver. These findings represent the use of distributed coherence in the V band without the use of GPS timing. 

\end{abstract}

\begin{IEEEkeywords}Beamforming, communications, distributed RF coherence, mmWave, optical synchronization, spatial-temporal signal processing
\end{IEEEkeywords}

%% file: sections/1_intro.tex
\section{Introduction}

Operating RF applications at millimeter-wave (mmWave) frequencies offers substantial advantages over lower frequencies, including significantly higher bandwidth and data rates. 
The shorter wavelengths of mmWave signals also allow for highly directional, tightly focused beams, improving spatial resolution for precise sensing and object tracking. 
The capabilities of mmWave systems can be expanded by using distributed, coherent elements. 

Distributed coherent systems can achieve substantial performance improvements over a single transceiver \cite{NanzerDist2021} in various applications, such as next-generation wireless communications, remote sensing, advanced radar networks, and deep-space exploration \cite{Jayaprakasam2017,Larsson2024,Mudumbai2007,Mudumbai2009,Holtom2024}. 
In particular, beamforming for communications signals can be implemented in a network of distributed, coherent platforms to achieve a gain in received power over the average \ac{SISO} link, bounded by the square of the number of platforms $N$.
Furthermore, such a system can achieve a large aperture beyond what can be feasibly achieved with just one platform. 
This attribute allows the system to precisely direct beams or generate nulls to mitigate interference.

The primary challenge involved in coordinating and using distributed platforms as an array lies in synchronizing time, frequency, and phase. This challenge is further exacerbated with higher carrier frequencies due to increased reference clock error, increased phase noise, reduced coherence time, and greater sensitivity to environmental perturbations. To achieve the required coherence stability to operate at 60 GHz, we integrate a previously developed, optical time synchronization network \cite{martin2023demonstrationrealtimeprecisionoptical}.

\begin{figure}[htb]
    \centering
    \subfloat[]{\label{fig:IntroRx}
    \includegraphics[width=0.75\linewidth]
    {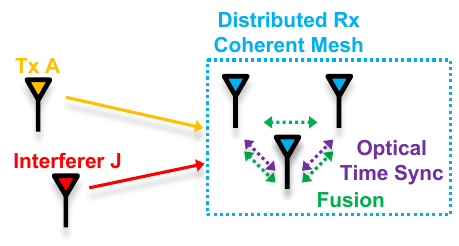}
    }
    
    \subfloat[]{\label{fig:IntroTx}
    \includegraphics[trim={0 0 0 0.5cm},clip,width=0.75\linewidth]
    {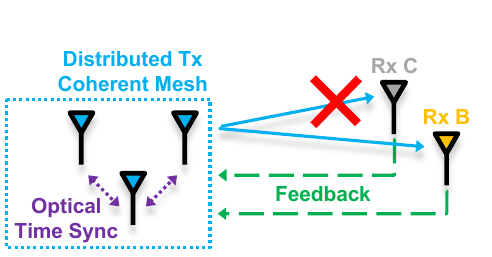}
    }
\caption{Multiobjective beamforming applications. (a) In distributed receive mesh beamforming, the mesh enhances the reception of a signal of interest while rejecting interference. (b) In distributed transmit beamforming, the mesh focuses its transmission toward one receiver while minimizing the power received at a second receiver. 
}
\label{fig:Intro}
\end{figure}

In this work, we study a set of distributed, coherent elements (referred to here collectively as a distributed coherent mesh network, where one node in the mesh is one single-antenna platform) for secure, robust, and efficient wireless communication. 
Specifically, we study a spatiotemporal beamforming mesh network operating under two main scenarios. 
The first scenario, shown in Fig.~\subref{fig:IntroRx}, involves receive beamforming with interference rejection, where an array of distributed receive mesh nodes is utilized to enable the constructive interference of a desired signal and simultaneously the deconstructive interference of a separate source (denoted as J). 
The second scenario, as shown in Fig.~\subref{fig:IntroTx}, involves transmit beamforming with nulling, in which an array of distributed transmit mesh nodes is employed to enhance the signal power at a primary receiver (denoted as Rx B) while also minimizing power directed at a second receiver (denoted as Rx C).

We previously demonstrated a distributed, coherent beamforming network with a slightly different network topology at sub-6 GHz \cite{WsicaNet2rd,holtom2023distributed}. The required coherence was sufficiently established through RF means while driving the RF platforms using standard crystal oscillators. 
In this work, we implement and demonstrate a 60 GHz over-the-air distributed multi-objective beamforming system without using GPS.

\subsection{Background}
The specific requirements on coherence precision depends on the application space. Beamforming applications can tolerate some amount of misalignment \cite{Mudumbai2007}. In terms of the application presented in Fig.~\ref{fig:Intro}, different challenges in maintaining the required coherence arise depending on the scenario. Transmit beamforming requires real-time coordination among the distributed nodes, in which the transmitted signals arrive at the receiver sufficiently time, frequency and phase-aligned. For receive beamforming, coherence can be achieved entirely through post-processing, which relaxes the precision of coherence required in real-time at the expense of computational complexity.
In contrast, applications, such as positioning, require even more precise alignment, otherwise ambiguous estimates arise. 

Several approaches can be leveraged to maintain coherence within distributed arrays. 
Closed-loop approaches rely on external feedback from a cooperative transceiver to construct corrective filters that achieve distributed coherence. In contrast, open-loop approaches use calibration techniques for clock synchronization \cite{Holtom2024,MerloTimeSync2023} and internode ranging \cite{mmWaveRanging2015,EllisonOpenLoopRanging2022} by constantly exchanging information among the mesh nodes without external input. Open-loop systems enable the additional ability to dynamically steer beams or scan as a distributed phased array from positional calibration without external feedback. Due to requiring precise positional knowledge of each node, this system is more challenging to implement compared to closed-loop systems. Most prior research has focused on closed-loop approaches \cite{ClosedLoop1_2002,ClosedLoop2_2008,ClosedLoop3_2012,ClosedLoop4_2012,ClosedLoop5_2012}, including retrodirective beamforming methods \cite{Retrodirective2010,Retrodirective2011}. 

Recent efforts in distributed beamforming, whether they utilize calibration techniques \cite{BhattacharyyaOpenLoopDist2024} or waveform exploitation and feedback provided by a transceiver external to the mesh \cite{WsicaNet2rd,holtom2023distributed}, primarily focus on sub-6 GHz applications. While these methods have proven effective at lower frequencies, they encounter limitations at higher frequencies, such as mmWave.  

\subsection{Contributions}

In this work, we demonstrate the ability to establish distributed coherence at mmWave frequencies allowing us to perform beamforming for robust communications. 

The major contributions of this work are summarized below: 
\begin{itemize}
    \item Establishment of mmWave distributed coherence by integrating \ac{OTS} into a network of distributed platforms, implementing an RF timestamp synchronization algorithm, employing fine frequency correction algorithms, and beamforming algorithms that provide fine-scale time and phase realignment; 
    \item Demonstration of distributed coherent transmit and receiving beamforming with interference nulling capability at 60 GHz using a three-element mesh; 
    \item Validation of the system’s robustness and stability in over-the-air demonstrations with interference.
\end{itemize}

\subsection{Notation}

Scalars are indicated by non-bold fonts, column vectors with bold lower case fonts, matrices with bold upper case fonts, row vectors with bold lower case fonts with an underline, transpose with $\cdot^\text{T}$, conjugate with $\cdot^*$, and
Hermitian conjugate (or conjugate transpose) with $\cdot^\text{H}$. The set of real variables are indicated by ${\mathbb R}$ and the set of complex variables by ${\mathbb C}$. An $\ell^2$-norm, or vector norm, is indicated by $\| \cdot \|$, and the Frobenius norm is given by $\| \cdot \|_F$. An estimate of a variable is indicated by $\hat{\cdot}$.

%% file: sections/2_sec.tex
\section{Theoretical Preliminaries}
 
To understand how to construct beamformers for each scenario, we first mathematically model the receptions. 
Traditional approaches can be used to derive the beamformers, but we modify their construction to make operation more robust. 
The derivations assume that the distributed mesh is time-frequency aligned. 
In practice, distributed elements are unlikely to be precisely phase aligned, so we present performance bounds as a function of synchronization errors to describe how performance degrades in consequence.

\subsection{Receive Beamforming}
In the receive beamforming scenario illustrated in Fig.~\subref{fig:IntroRx}, the mesh aims to improve the \ac{SNR} of a desired transmitted signal while mitigating any interference. 
Each array node receives a signal that is a combination of the desired signal $s[t]$ emitted from the transmitter (labeled Tx A) and an undesired signal $j[t]$ from an interferer (labeled interferer-J). Desired signal $s[t]$ propagates through a channel modeled as the finite impulse response filter, $h_{A,n}[t]$, to arrive at mesh node $n$. Similarly, interfering signal $j[t]$ travels through a corresponding channel $h_{J,n}[t]$ to arrive at node $n$. 
Here we assume the presence of a single interferer, but mitigating multiple can be supported provided there are sufficient \ac{DoF}. 
Each mesh node receives a signal, 
\begin{align}
    z_n[t] = (h_{A,n} * s)[t-\tau_A] + (h_{J, n} * j)[t-\tau_j] + q_{n}[t] \, ,
    \label{eq:rxbf_rxsig}
\end{align}
where $\tau_A$ and $\tau_j$ are the respective \acp{ToF} from the source and jammer to their respective closest mesh node. The \acp{TDoA} are all characterized within the channel impulse responses. The combined effect of time-shifting by $\tau_A$ and convolving by $h_{n,A}[t]$ shifts the arrival of $s[t]$ at each node to the correct time. We assume the noise signal $q_{n}[t]$ is an additive white complex Gaussian noise process, independent across nodes.

The effects of beamforming are achieved virtually by post-processing the signals. The $N$ mesh nodes offload their receptions to a centralized processing location. This post-processor applies to $z_n[t]$ a corresponding beamforming filter, $w_n^*[t]$, of length $T_w$ taps and sums the outputs, yielding the beamformed reception, 
\begin{align}
    x[t]&= \sum_{n=1}^N  (w_n^* * z_n)[t]
    \,.
    \label{eq:rxbf_compsig}
\end{align}

To improve the \ac{SINR} of the desired signal, we derive a spatiotemporal beamformer (the collection of filters $w_n[t]$ $n\in\{1, \dots,N\}$) by minimizing the mean squared error between the resulting post-processed receptions and the training information embedded within $s[t]$.

To better convey the beamformer's construction, we can alternatively express the convolution and sum in (\ref{eq:rxbf_compsig}) as a vector-matrix product. 
We first identify a $T_z$-sample window of $z_n[t]$ starting at time $\hat{\tau}$, the estimated earliest arrival of $s[t]$. 
We construct a data-delay matrix $\tilde{\textbf{Z}}_n \in \mathbb{C} ^{T_w \times (T_z+T_w-1)}$, defined as 
\begin{align}
    \tilde{\textbf{Z}}_n = 
    \begin{bmatrix}
        z_n[\hat{\tau}] & \dots  & \!z_n[\hat{\tau}\!+\!T_z\!-\!1]\! & \dots  & 0 \\
        \\
              \vdots    & \ddots &                 & \ddots & \vdots\\
        \\
              0         & \dots  & z_n[\hat{\tau}]       & \dots & \!z_n[\hat{\tau}\!+\!T_z\!-\!1]\!
    \end{bmatrix}
    \label{eq:spatiotemporal}
    \,.
\end{align}
We define a vector $\underline{\bf w}_n = \begin{bmatrix}
    w_n[0] & \dots & w_n[T_w-1]
\end{bmatrix}$, containing the beamforming filter coefficients prescribed to node $n$. 
Further, we form a space-delay data matrix $\tilde{\textbf{Z}} \in \mathbb{C} ^{NT_w \times (T_z+T_w-1)} $ by stacking all $\tilde{\textbf{Z}}_n$ matrices together according to the structure, $\tilde{\textbf{Z}}=\begin{bmatrix}
    \tilde{\textbf{Z}}_1^\text{T}&
    \dots&
    \tilde{\textbf{Z}}_N^\text{T}
    \end{bmatrix}^\text{T}$.
We define the vector $\textbf{w}=\begin{bmatrix}
        \underline{\bf w}_1&
        \hdots&
        \underline{\bf w}_{N}
    \end{bmatrix}^{\text{T}}$, 
which contains all $N$ beamforming filters. Evaluating ${\textbf{w}}^\text{H}\tilde{\textbf{Z}}$ accomplishes the convolutions and the summation.
Lastly, we define the training data vector $\underline{\bf s} \in \mathbb{C}^{1\times (T_z+T_w-1)}$, which contains $T_z$ samples of training information with zero-padding at the beginning to ensure that the the energy of the filters is centered within the $T_w$ taps and zero-padding at the end to resolve sizing mismatches. 

The optimization problem is given by 
\begin{align}
    \underset{\textbf{w}}{\min} \quad \left\| {\textbf{w}}^\text{H}\tilde{\textbf{Z}} - \underline{{\mathbf s}} \right\|^2 
    \,,
    \label{eq:OptProb}
\end{align}
which has the solution, $\left(\tilde{\textbf{Z}}\,\tilde{\textbf{Z}}^\text{H}\right)^{-1} \, \tilde{\textbf{Z}}\,\underline{\bf s}^\text{H}$.

The covariance term $\tilde{\textbf{Z}}\,\tilde{\textbf{Z}}^\text{H}$ can present several practical issues, including a poor condition number and insufficient integration. There are several techniques to alleviate these issues, such as matrix factorization, tapering, or L$_2$ regularization. Here we utilize L$_2$ regularization, or diagonal loading, to avoid inverting a matrix that is close to being singular. A tractable solution can be obtained by instead computing 
\begin{align}
\textbf{w} =
\left(\tilde{\textbf{Z}}\,\tilde{\textbf{Z}}^\text{H} 
+ \delta{\bf I}\right)^{-1} \, \tilde{\textbf{Z}}\,\underline{\bf s}^\text{H}
\,,
\label{eq:MMSEWideband}
\end{align}
where $\delta$ is a small factor. Furthermore, $\tilde{\textbf{Z}}\,\tilde{\textbf{Z}}^\text{H}$ contains both a spatiotemporal covariance structure attributed to the channels $h_{A,n}[t]$ as well as $h_{J,n}[t]$. Interference rejection capabilities are provided by inverting $\tilde{\textbf{Z}}\,\tilde{\textbf{Z}}^\text{H}$. If this covariance matrix is at risk of being inaccurately estimated, which may naturally happen due to systematic or environmental dynamics, then there is a risk of self-nulling. To address this, we can instead compute $\tilde{\textbf{Z}}\,\tilde{\textbf{Z}}^\text{H}$ using data that contains strictly interference and noise, thus leaving only noise and $h_{J,n}[t]$ spatiotemporal covariance information. We can obtain this information through an ambient observation period. Leveraging this modification, however, may not always be available to the system. For example, the nature of the communications protocol may not allow sufficient observations. Moreover, diagonal loading also contributes to self-nulling and can reduce interference rejection performance. The value of $\delta$ is chosen to mitigate these negative effects.  

\subsection{Transmit Beamforming} \label{sec:BeamFormer}

In the transmit nulling scenario illustrated in Fig.~\subref{fig:IntroTx}, the mesh focuses transmitted energy toward a target receiver, while concurrently minimizing the energy directed toward another receiver. 
We explore two variations of transmit beamforming. The first involves strictly maximizing the \ac{SNR} of the desired signal arriving at a single receiver. The other includes the additional challenge of minimizing the received energy of the transmitted signal at a second receiver. 

To understand how to construct a transmit beamformer, we first develop models of the signals received at both receivers from the transmitting mesh nodes. 

Receiver B and Receiver C observe a joint transmission from the distributed mesh.
Mesh node $n$ applies a beamforming filter $w_n^*[t]$ to the common signal $s[t]$ and transmits $(w_n^* * s)[t]$. The predistorted signals emitted by the mesh nodes propagate through corresponding channels $h_{n,B}[t]$ $n\in\{1, \dots, N\}$ and combine upon arrival, such that Receiver B receives the composite signal, 
\begin{align}
    g[t]&= \sum_{n=1}^N  (h_{n,B} * w_n^* * s)[t-\tau_B] + q_B[t] \, ,
    \label{eq:txbf_rxsig_prim}
\end{align}
where $\tau_B$ is the \ac{ToF} from the shortest path between the mesh node and Receiver B. Again, the \acp{TDoA} are all characterized within the channel impulse responses. We assume the noise signal $q_B[t]$ is an additive white complex Gaussian noise process. 
Receiver C receives a signal defined analogously to (\ref{eq:txbf_rxsig_prim}), 
\begin{align}
    r[t]&= \sum_{n=1}^N  (h_{n,C} * w_n^* * s)[t-\tau_C] + q_C[t] \, ,
    \label{eq:txbf_rxsig_sec}
\end{align}
where $h_{n,C}[t]$ are the channels bridging mesh nodes to Receiver C, $\tau_C$ is the shortest \ac{ToF}, and $q_C[t]$ is an additive white complex Gaussian noise process.

In the first described variation, assuming a dispersive channel, one method we can use to maximize the receive \ac{SNR} of a single source's signal is by deploying a spatiotemporal matched filter beamformer. 
 
In this case, we cannot post-process a composite reception to achieve beamforming. Instead, we must pre-distort each transmission with a corresponding filter that achieves matched filtering for each transmission upon arrival. The signals arrive coherently combined, resulting in improved \ac{SNR}.
Constructing and applying effective pre-distortion requires knowledge of the channel beforehand. 
We choose to have receivers obtain estimates of $h_{n,B}[t]$, denoted as $\hat{h}_{n,B}[t]$, which should characterize the \acp{TDoA}, and feed them back to the mesh. We assume a tapped delay line model for the channel and estimate the taps using least squares \cite{Kay1993Vol1Est}. The efficacy of this beamformer is predicated on the channel coherence interval at the time of transmission to allow for a valid estimate of $h_{n,B}[t]$.

After being provided the relevant channel estimates, mesh nodes perform $N$ independent optimizations to produce the beamforming filters. Each node solves the optimization problem, 
\begin{align}
\underset{w_n[t]}{\max} &\quad \frac{\text{E}\left[\left\vert(\hat{h}_{n,B} * {w}_n^* * {s})[t]\right\vert^2\right]} {\text{E}[\left\vert q_B[t]] \right\vert^2]} \\
\text{subject to}  &\quad \left\Vert{w}_n[t]\right\Vert^2 = 1 \,.
\end{align}

When treating this problem in the frequency domain, setting $W^*(-f) = \hat{H}^*(f)/ \sqrt{\int_{-1/2}^{1/2} |H^*(f)|^2 df}$ maximizes the \ac{SNR}, where the normalization term causes the solution to satisfy the power constraint. 
The resulting impulse response of the matched filter is given by $w_n[t]= \hat{h}_{n,B}[-t]/\Vert \hat{h}_{n,B}[-t]\Vert$, where the complex conjugate applied to $w_n[t]$ in (\ref{eq:txbf_rxsig_prim}) is crucial for realigning phase. All nodes should apply a common timing offset across to make each $w_n[t]$ causal.

If an estimated channel, of length $T_h$ is stored in vectors given by $\underline{\hat{\mathbf{h}}}_{n,B}= \begin{bmatrix} \hat{h}_{n,B}[0], \dots,\hat{h}_{n,B}[T_h-1] \end{bmatrix}$, then the corresponding matched filter beamformer is 
\begin{align}
        \underline{\bf{w}}_n = \frac{1}{\Vert\underline{\hat{\bf{h}}}_{n, B}\Vert_2}\,\begin{bmatrix}\hat{h}_{n,B}[T_h-1]& \hdots& 
        \hat{h}_{n,B}[0]\end{bmatrix}\,.
        \label{eq:STMF}
\end{align}

In the second variation, transmit nulling, the additional challenge of minimizing the received energy of a transmitted signal at a second receiver requires a different approach. We use a modified minimum mean squared error construction facilitated by feedback in the form of channel estimates from both Receiver C and Receiver B to construct a null in the direction of Receiver C and direct energy towards Receiver B. 

We consider a narrowband signal model for now, so the impulse response of each channel is represented by just one complex-valued scalar. We define vectors containing the estimates of each channel corresponding to Receiver B and Receiver C as,  ${\hat{\mathbf{h}}_B}  = \begin{bmatrix}
\hat{h}_{1,B}, \hat{h}_{2,B},\dots, \hat{h}_{N,B}\end{bmatrix}^\text{T}$ and ${\hat{\mathbf{h}}_C} = \begin{bmatrix} \hat{h}_{1,C},\hat{h}_{2,C},\dots,\hat{h}_{N,C}\end{bmatrix}^\text{T}$ respectively. 

We define the set of beamforming weights as an $N$-element column vector,  ${\bf w}$. The result of ${{\mathbf{h}_B}^\text{T}}
{\bf w}^* \underline{\bf s}$ and ${\bf h}_{C}^\text{T}{\bf w}^* \underline{\bf s}$ are hypotheses of the beamformed signals impinging on Receiver B and C respectively. We jointly optimize ${\bf w}$ such that ${{\mathbf{h}_B}^\text{T}}
{\bf w}^* \underline{\bf s}$ matches $\underline{\bf s}$ and ${\bf h}_{C}^\text{T}{\bf w}^* \underline{\bf s}$ matches $\underline{\bf 0}$ as close as possible. Performing the minimization

\begin{align}
    \underset{{\bf w}}{\min} \quad &\left\Vert \begin{bmatrix} {\bf h}_{B}^\text{T}
{\bf w}^* \underline{\bf s} \\ {\bf h}_{C}^\text{T}{\bf w}^* \underline{\bf s} \end{bmatrix} - \begin{bmatrix} \underline{\bf s} \\ \underline{\bf 0} \end{bmatrix}\right\Vert^2_F\,,
\end{align}
yields the solution, ${\bf w} =  ( \hat{\bf h}_{B} \hat{\bf h}_{B}^\text{H} + \hat{\bf h}_{C} \hat{\bf h}_{C}^\text{H} )^{-1} \hat{\bf h}_{B}\,.$

This solution is modified to address rank deficiency and self-nulling. We can again apply diagonal loading to permit an inversion and regularize the result. 
Similar to the MMSE receive beamforming solution, the presence of the term $\hat{\bf h}_{B} \hat{\bf h}_{B}^\text{H}$ in the covariance can potentially hinder increasing received power toward the Receiver B. 

We exclude $\hat{\bf h}_{B} \hat{\bf h}_{B}^\text{H}$ from the covariance, while retaining our intended nulling capabilities facilitated by the $\hat{\bf h}_{C} \hat{\bf h}_{C}^\text{H}$ term. 
With both modifications, the implemented beamformer weights become
\begin{align}
    {\bf w} = & \left(  {\hat{\bf h}}_{C} {\hat{\bf h}}_{C}^\text{H}  + \delta \mathbf{I} \right)^{-1} {\hat{\bf h}}_{B} \,.
    \label{eq::Trans_Null_MMSE_Mod}
\end{align}
       
\subsection{Performance Bounds} \label{Perf_Met_Bounds}

A requirement for successful distributed beamforming is time-frequency synchronization within the mesh. There are differences in the true clock frequencies driving each node in the mesh. Furthermore, the physical system is dynamic, meaning the offsets between all nodes are constantly changing. Residual offsets are inevitable because we cannot instantaneously estimate and apply corrections. We update correction estimates regularly at a rate to limit the phase offset accumulated over time. The performance bounds for power gain and \ac{INR} reduction are a function of total accumulated phase error variance from all sources of error.

With some modification to the work found in \cite{Bakr09, Bakr09AmpError, Holtom2024}, a bound on power gain, assuming the SISO SNRs are the same for every link, can be expressed as a function of this accumulated phase error variance, $\varphi^2$:
    \begin{align}
        \text {Power Gain} \leq(N^2-N)\left(e^{-\varphi^{2}}\right) + N
        \,.
    \end{align}

In transmit beamforming, the \ac{SNR} gain bound is equivalent to the power gain bound. 

However, due to the post-processing required in receive beamforming, $N$ independent noise signals are combined in addition to the $N$ receptions of the signal of interest. Even though the power gain bound still holds, the noise power increases by a factor of $N$, resulting in a reduced bound on the SNR gain. 
Without accumulated phase error variance, these bounds result in a maximum \ac{SNR} gain of $N^2$ in transmit beamforming experiments and $N$ in receive beamforming experiments. 
Likewise, with some modification to the work in \cite{Holtom2024}, the bound on INR reduction, assuming equal SISO INRs across all links, can also be expressed as a function of $\varphi^2$, and is given as
    \begin{align}
        \text {\ac{INR} Reduction} \leq\frac{N-1}{N} \left(1 - e^{-\varphi^{2}}\right)
        \,.
    \end{align}

Several sources contribute to the variance in phase offset, such as CFO estimation error (scaled by an estimate update rate), timing jitter, and phase offset estimation error (due to building the beamformers). In a real-time environment, there will be more accumulated errors than one can estimate. Therefore, we aim to reduce this variance to reach the bound by decreasing the update interval and precisely estimating time and frequency errors that contribute to the accumulated phase error.
  
Achieving \ac{SNR} gain tends to be more tolerant to errors, and achieving effective \ac{INR} reduction is much less tolerant. The adage that beams are wide and nulls are narrow holds in the context of distributed coherent beamforming. These theoretical bounds on \ac{SNR} gain and \ac{INR} reduction allow one to determine a tolerable amount of accumulated phase error to meet desired performance.

%% file: sections/3_sec.tex
\section{Experimental Beamforming System Description}

\begin{figure}[!b]
        \centering        \includegraphics{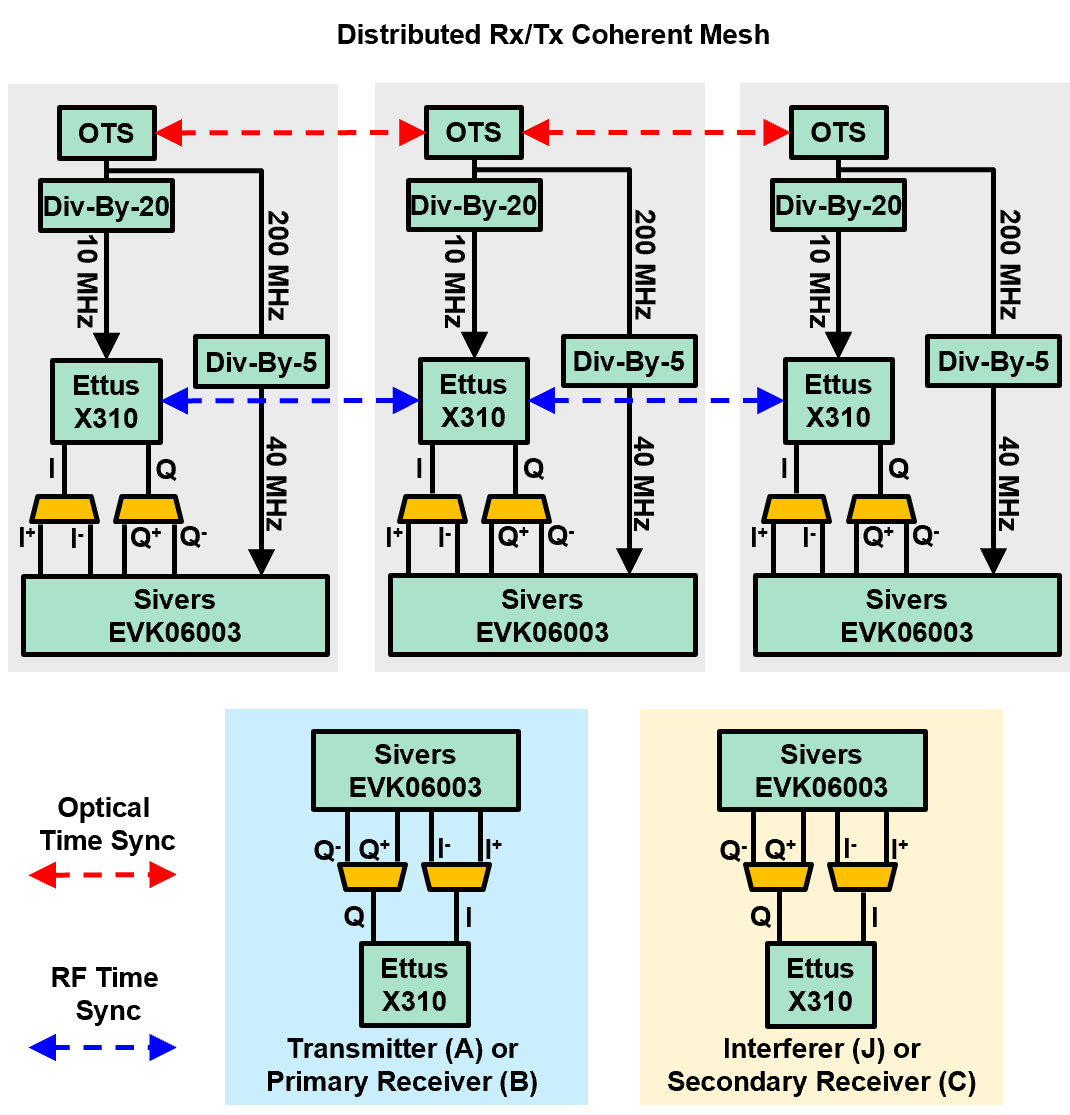}
        \caption{System-level block diagram showing component interface and clock distribution for both beamforming applications.
        }
        \label{fig:Rx_Clock}
    \end{figure}

We constructed an experimental setup using software-defined radios to demonstrate transmit and receive beamforming for the scenarios depicted in Fig. \ref{fig:Intro}. The interfacing of all the radios to produce each configuration is illustrated in Fig. \ref{fig:Rx_Clock}. 
The core of each RF platform involved in the experiments is an Ettus USRP X310 software-defined radio paired with a Sivers mmWave frontend. 
The X310s have two transceiving channels available for use. 
Mesh nodes are configured to utilize both.
The first channel is equipped with Ettus BasicRX and BasicTX daughterboards, which feed a mmWave front-end module from Sivers-EVK6003. The mmWave RF front-end delivers to and accepts from the X310 differential analog baseband, zero-IF I/Q signals. These signals are exchanged through four TX or RX micro-miniature coaxial (MMCX) connectors.
The other channel is equipped with an Ettus UBX-160 daughterboard, connected to a 3 dBi dipole antenna rated for 900 MHz. This channel facilitates an RF side link for timestamp synchronization.  
Other transmit, receive, and interferer nodes outside of the mesh use an X310 outfitted strictly with a pair of BasicRX and BasicTX daughterboards that are attached to additional SIVERS mmWave front-ends. 

The OTS and RF time transfer sub-systems are crucial for achieving time-frequency synchronization between mesh nodes. 
Frequency and time alignment between the three OTS nodes is achieved through optical time synchronization, depicted by the red two-way arrow in Fig. \ref{fig:Rx_Clock}. 
Additionally, time-stamp synchronization is established through the dedicated 900 MHz RF channel, depicted by the blue two-way arrow in Fig. \ref{fig:Rx_Clock}.

The X310s utilize a 10 MHz clock source, while the mmWave front-ends require a 40 MHz reference. 
The optical sub-system can provide clock signals at two frequencies, 200 MHz, the fundamental, and 10 MHz, which is derived. 
The X310 can directly use the 10 MHz clock signal through the external clock reference port.  
However, the 200 MHz clock signal must be passed through a frequency divider (Lotus Communication Systems, Model No. FD5DC7G), which outputs a 40 MHz clock signal suitable for the mmWave front-end's MMCX connector. 

\begin{figure}[!t]
        \centering
        \includegraphics[width=\linewidth]{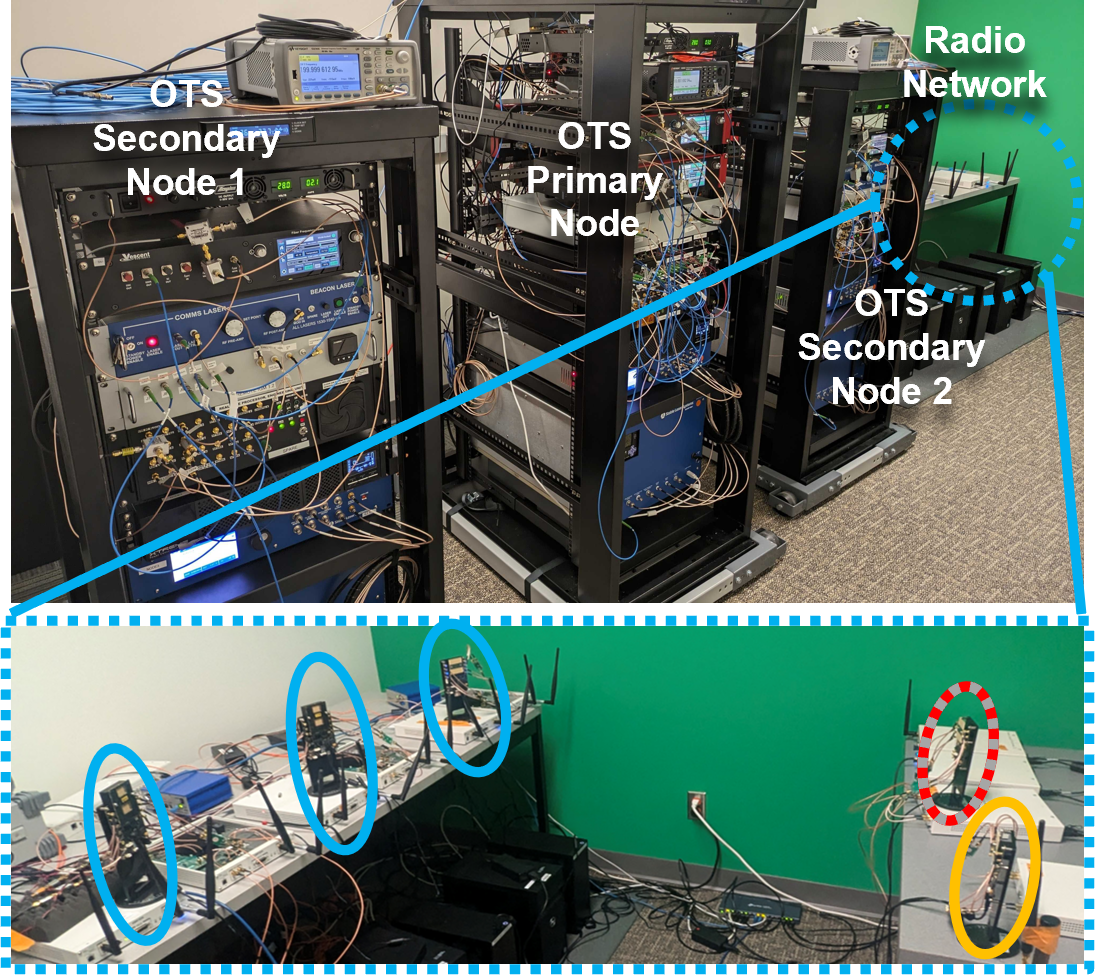}
        \caption{The overall system setup includes a three-node OTS system and an mmWave radio system. Each radio sub-system is comprised of a baseband SDR interfacing with a mmWave frontend.}
        \label{fig::System}
    \end{figure}

The whole experimental setup is depicted in Fig. \ref{fig::System}.

\subsection{Optical Time Synchronization} \label{Sec::OTT}
\begin{figure}[!ht]
        \centering
        \includegraphics[width=\linewidth]{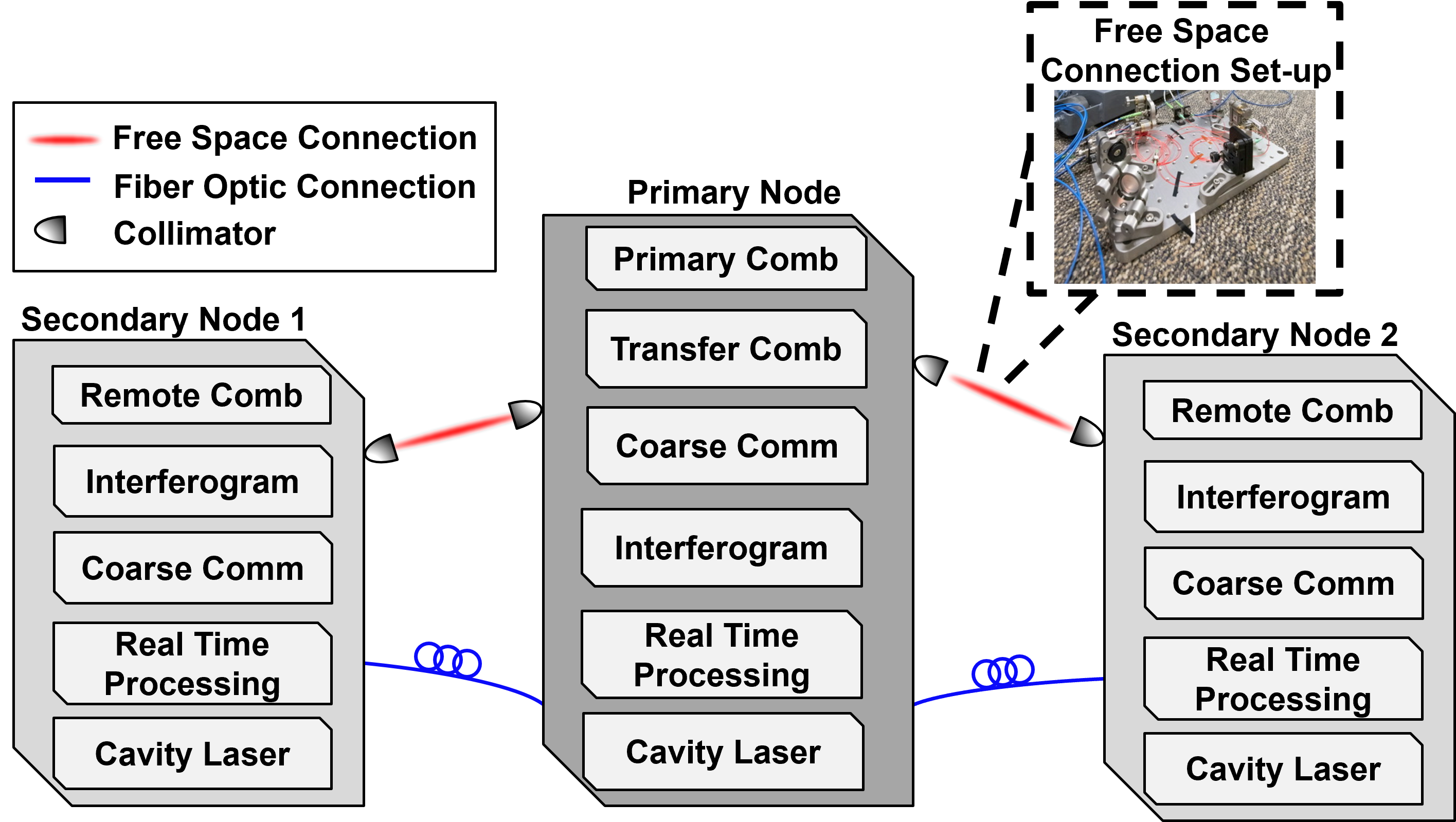}
        \caption{A simplified representation of the internal components and connections of the \ac{OTS} system. In the default configuration, the connections between the primary node to secondary nodes are through fiber optic cables. In an additional configuration, these connections are made through free space by using the depicted light table.} 
        \label{fig::OTT-config}
    \end{figure}

    Air Force Research Laboratory \cite{martin2023demonstrationrealtimeprecisionoptical} developed the \ac{OTS} system utilized in this work to achieve femtosecond-level timing precision across multiple distributed nodes. 
    This sub-system provides a method for aligning independent clock references in a distributed manner. 
    The system implements a star topology, where a central primary node distributes time signals to secondary nodes via fiber optic cables or free-space over-the-air laser communication, as shown in Fig. \ref{fig::OTT-config}. Each node contains an independent cavity-stabilized laser and frequency comb(s), which serve as the physical oscillators. The system synchronizes frequency combs through a two-way protocol that uses a combination of linear-optical-sampling (LOS) and phase-modulated optical signals \cite{jd2016prx} to establish time synchronization. A feedback loop that adjusts the clocks at the secondary nodes to align with the primary node to maintain synchronization.

    The two-way LOS process involves the exchange of frequency comb pulses between each pair of nodes (i.e., the primary node and the respective remote nodes). Each respective node optically mixes these pulses to generate interferograms, which encode the timing information. By measuring the arrival times of these interferograms and exchanging their respective timestamps over the optical links, the system calculates the clock offsets and adjusts the secondary clocks in real-time. This process ensures high precision and stability, enabling time synchronization with accuracy on the order of 1 fs or less. This level of synchronization represents a significant leap in distributed timing and enables applications in mmWave coherence systems.

    \subsection{RF Time Transfer for Timestamp Synchronization} \label{Sec::RF}
        
            Although the \ac{OTS} addresses relative and absolute time offsets, this work only utilized the clock output from each node, without additional timestamping information.
            Establishing a common time origin among the coherent mesh helps facilitate beamforming. 
            To accomplish this goal, we implement an over-the-air RF time transfer sub-system, consisting of a \acf{NTP} used for timestamp synchronization within the distributed beamforming mesh. 
            A version of a leader-and-follower-based \acs{NTP} was implemented. One node is designated the leader, labeled as node $L$. All other nodes are followers labeled $n \in \mathcal{S}$, where $\mathcal{S}$ contains all follower node labels and $L\notin\mathcal{S}$. Followers in the mesh adjust timing to align with the leader.

            \begin{figure}[!ht]
                \centering
                \includegraphics[width=0.6\linewidth,trim={0 3cm 0 3cm},clip]{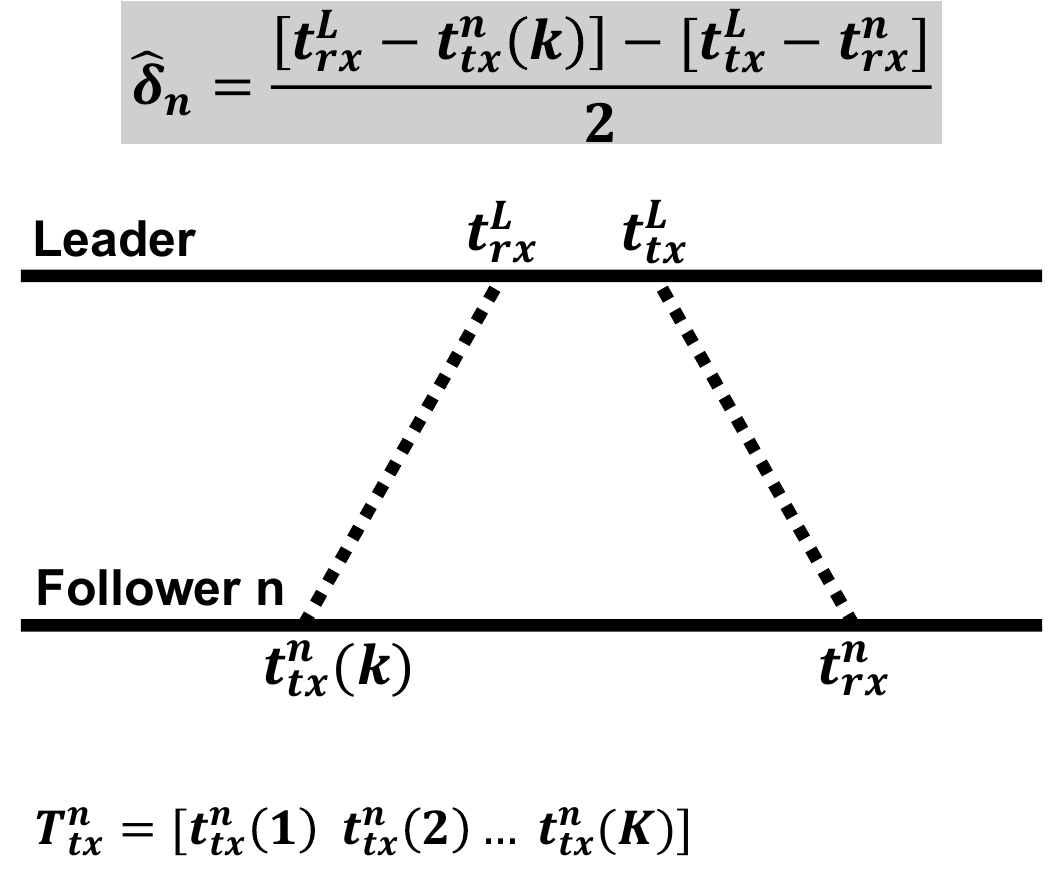}
                \caption{The leader-follower-based \acs{NTP} approach used to synchronize local clock sources.}
                \label{fig:timesynch}
            \end{figure}
 
                Consider that the leader's timestamps are ahead of follower node $n$'s timestamps by offset $\delta_n$. 
                Follower node $n$ transmits a message containing its local transmission timestamp, $t_{tx}^n$. 
                The leader, node $L$, acquires the signal and records the estimated \ac{ToA}, $t_{rx}^L$, with respect to its time axis. The leader also unpacks $t_{tx}^n$ from the reception. The leader then transmits a message containing its local transmission timestamp, $t_{tx}^L$, the reception timestamp, $t_{rx}^L$, and the unpacked timestamp, $t_{tx}^n$, back to the same follower.
                The follower receives the message, performs acquisition, and records its local reception time, $t_{rx}^n$. This process is illustrated in Fig. \ref{fig:timesynch}.
                
                Assuming the \ac{ToF} does not change significantly between transmission stages, clock offset $\delta_n$ between follower $n$ and the leader $L$ is estimated by evaluating 
                \begin{align}
                    \hat{\delta}_n = \frac{1}{2} \left(\left(t_{rx}^L - t_{tx}^n\right) - \left(t_{rx}^n - t_{tx}^L\right)\right)\,.\label{eq:RFtimetransfer}
                \end{align}
                We align the time axes by adjusting the scheduling of transmission and reception tasks. Specifically, we add the estimated offset $\hat{\delta}_n$ to node $n$'s local timestamps.

                To lighten the communications traffic, an index $k$, such that $t_{tx}^n= t^n[k]$, where $t^n[k]$ is a sequence of time stamps, can be transmitted instead of $t_{tx}^n$ explicitly. 
                If the follower holds in memory a short sequence of $t^n[k]$, then the memory required to transmit this reference can be reduced. 
                The leader node cannot encode the timestamps in this manner and must send back $t_{tx}^L$ and $t_{rx}^L$ with as much resolution as possible. We practically implement the over-the-air RF time transfer sub-system using the parameters summarized in Table \ref{Table::TS_Parameters}. 
                We perform the calculation of (\ref{eq:RFtimetransfer}) on the integer components and the fractional components of the time stamps separately, allowing us to maintain 64 bits of fractional precision on $\hat{\delta}_n$.

                \begin{table}[!ht]
    \centering
    \caption{RF TIME TRANSFER PARAMETERS}
    \begin{tabular}{ c c }
        \toprule[1pt]
            Parameter & Value \\
        \midrule
            Time Transfer Sampling Rate (${f_s}$)        &  2 MHz\\
            Time Transfer Bandwidth (${B}$)        &  1 MHz\\
            Time Transfer Carrier Frequency (${f_c}$)        &  902 MHz\\
            Modulation Scheme       &  QPSK\\
            Inner FEC     &  Golay\\
            Outer FEC     &  Hamming\\
            Time Transfer Preamble Length     &  512 samples \\
            Time Transfer Integer and Fractional Bit Depth    &  64 bits \\
        \bottomrule[1pt]
    \end{tabular}
    \label{Table::TS_Parameters}
\end{table}

\subsection{Experiment Description}
In the variants of beamforming experiments discussed, beamforming weights must be updated regularly to adapt to environmental and systematic dynamics, primarily time-varying channels. To adapt to this situation, we transmit frames in burst intervals, where each frame contains a preamble through which the beamforming mesh can update the beamformers. The burst intervals are also referred to as cycles.

\subsubsection{Receive Beamforming}

For the receive beamforming experiment, the transmitter transmits bursts. Meanwhile, an interferer is constantly transmitting. In this variation of receive beamforming, we consider a protocol such that all of the data can be fused, or pooled in a common processing location, and that there is time for the receiver to optimize beamformers for each frame and apply them. The information is post-processed after transmitting both the source and interferer, converting to baseband, and saving out the raw in-phase and quadrature components of the data from each node at each cycle.

\subsubsection{Transmit Beamforming}

For transmit beamforming, mesh nodes must estimate beamformers in real time, so that when they are applied, the effects of beamforming are achieved upon reception at the receiver or receivers. 
The beamforming mesh transmits a predistorted waveform every transmission cycle with the training information untouched. The receivers must estimate the channel and then return it to the mesh. The beamforming mesh uses these estimates to update the beamformer. Due to the feedback required to construct the beamformers using the methods described in Section \ref{sec:BeamFormer}, granted accurate channel estimates, the beamformer is most optimal given the channel state at the reception time. The dynamics of the system and environment may naturally cause some mismatch between the estimated beamformer and the optimal beamformer at the time of the next reception. How well the estimated beamformer fits this new scenario depends on the system's stability and the environment, including the relative positioning of all receiver-transmitter pairs.

\subsubsection{Key System Parameters}
    
    The implemented system operates at 60.48 GHz with a waveform bandwidth of 1 MHz and an RF power of approximately 100 mW. The duration of experiments varies between 30 seconds and 4 minutes, and transmit bursts are scheduled every 200 ms or 250 ms, depending on the experimental configuration. The system also utilizes a side channel within the 900 MHz frequency band dedicated to timestamp synchronization among distributed nodes, ensuring precise coordination. The key system parameters are summarized in Table \ref{Table::Parameters}.
    
\begin{table}[!ht]
    \centering
    \caption{EXPERIMENTAL SYSTEM PARAMETERS}
    \begin{tabular}{ c c }
        \toprule[1pt]
            Parameter & Value \\
        \midrule
            Sampling Rate (${f_s}$)        &  2 MHz\\
            Bandwidth (${B}$)        &  1 MHz\\
            Carrier Frequency (${f_c}$)        &  60.484 GHz\\
            Amble and Payload Modulation Scheme       &  QPSK\\
            Amble and Payload Length    &  8192 samples \\
            Estimation Integration Length   &  2048 samples \\
            Cycle Rate        &  200/250 ms\\
            Mesh Population ($N$)        &  3 nodes \\
        \bottomrule[1pt]
    \end{tabular}
    \label{Table::Parameters}
\end{table}

\subsection{Waveform Design}    

            We designed the waveforms transmitted during our experiments to facilitate performance evaluation and to ensure accurate estimates of key correction parameters, such as \ac{CFO} and sample time delay. 

            We generate preambles and postambles using pseudorandom binary sequences, specifically maximum length sequences (MLSs), modulated with \ac{QPSK}. These sequences have an autocorrelation that facilitates accurate parameter estimation, particularly for estimates of any filter. 
            The data payloads are all random \ac{QPSK} sequences. Guard intervals, lasting 256 samples, separate each section of the waveforms.
 
            Unique to the receive beamforming experiment, we introduce an interferer to emulate a realistic environment. 
            The interference waveform is a 256-\ac{QAM} sequence. 
            
\begin{figure}[!ht]
            \centering\includegraphics[width=0.8\linewidth]{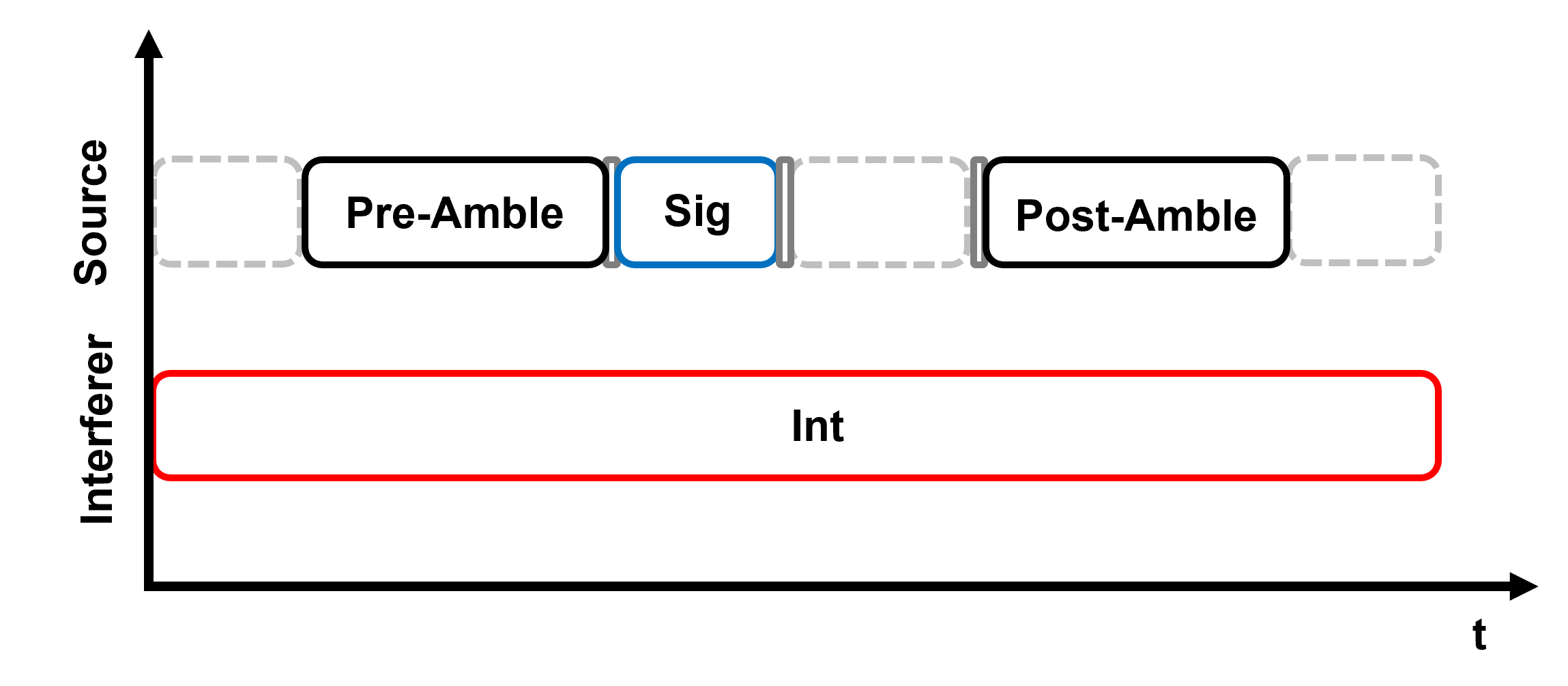}
            \caption{Waveforms transmitted during receive beamforming experiments. A ``look-through period" is intentionally placed after the signal of interest, designated ``Sig", to assist with assessing the interference. This lends to calculating the gain metrics of interest.}
            \label{fig:rxbf_wave}
            \end{figure}

            The waveforms transmitted by the source and interferer in the receive beamforming experiments are illustrated in Fig. \ref{fig:rxbf_wave}. These transmissions are of length 75560 samples. Aside from the preamble, data payload (designated ``Sig"), post-amble, and guard intervals, one frame of the source's transmission includes a period of silence, referred to here as a ``look-through" period. 
            The larger, dashed and gray rectangle in Fig. \ref{fig:rxbf_wave} following the source payload indicates this look-through interval. 
            The interferer continuously transmits alongside the source radio's transmission. The look-through assists in the assessment of inference power. Moreover, these samples are used to construct the spatiotemporal covariance matrix necessary to compute the MMSE Beamformer. 
            
\begin{figure}[!ht]
            \centering
            \includegraphics[width=\linewidth]{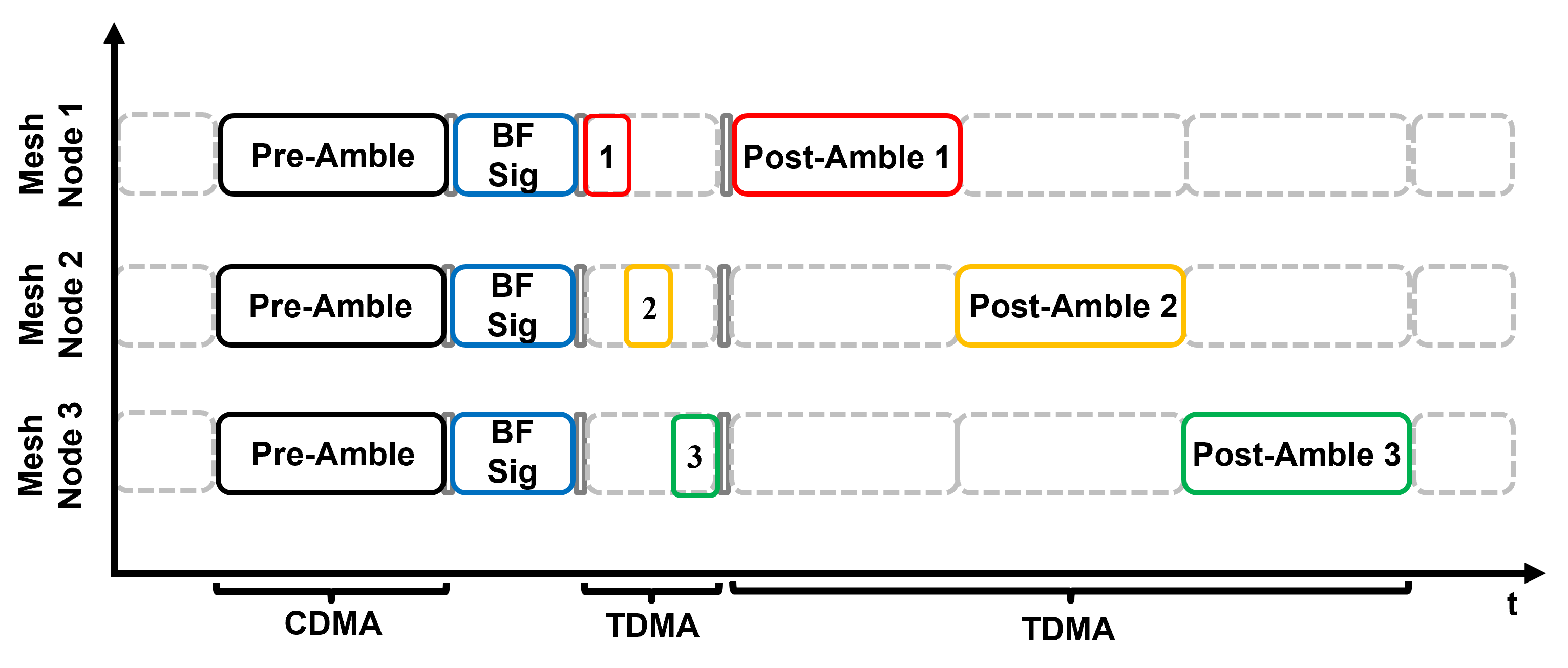}
            \caption{Waveforms deployed in transmit beamforming experiments. 
            All waveforms contain a preamble, and the signal of interest is predistorted to achieve beamforming. Using a \ac{TDMA} protocol, each node also transmits a signal for performance monitoring and a postamble, indicated by the colored sections.
            }
            \label{fig:txbf_wave}
            \end{figure}

            The waveforms transmitted by each mesh node during the transmit beamforming experiments are illustrated in Fig. \ref{fig:txbf_wave}. These transmissions are longer than those in receive beamforming, with 91472 samples to accommodate the \ac{TDMA} of the post-amble and individual \ac{SISO} transmissions.
            The nodes transmit orthogonal pre-ambles concurrently, in a \ac{CDMA} strategy. Assuming accurate timestamp synchronization, the preamble provides a common reference point for parsing the received composite signal. Additionally, all channel states can be estimated at the same time. The nodes also simultaneously transmit predistorted payload signals, labeled as ``BF Sig", to achieve beamforming. 

            Some beamforming performance metrics are measured relative to the performance of an average \ac{SISO} link, so we insert additional sections into the waveform, where each node transmits a segment of the non-beamformed payload in a predetermined \ac{TDMA} format.  
            These transmissions are the colored, numbered sections in Fig. \ref{fig:txbf_wave}.
            Each reception of the single-input transmission permits a power estimate from the corresponding link. 
            Additionally, nodes transmit a postamble using \ac{TDMA} to allow the receiver to estimate key parameters without interference from other nodes. 

%% file: sections/4_sec.tex
\section{Baseband Processing}
    \subsection{Receive Beamforming}

            \begin{figure}[!ht]
        	\centering
        	\includegraphics[width=\linewidth]{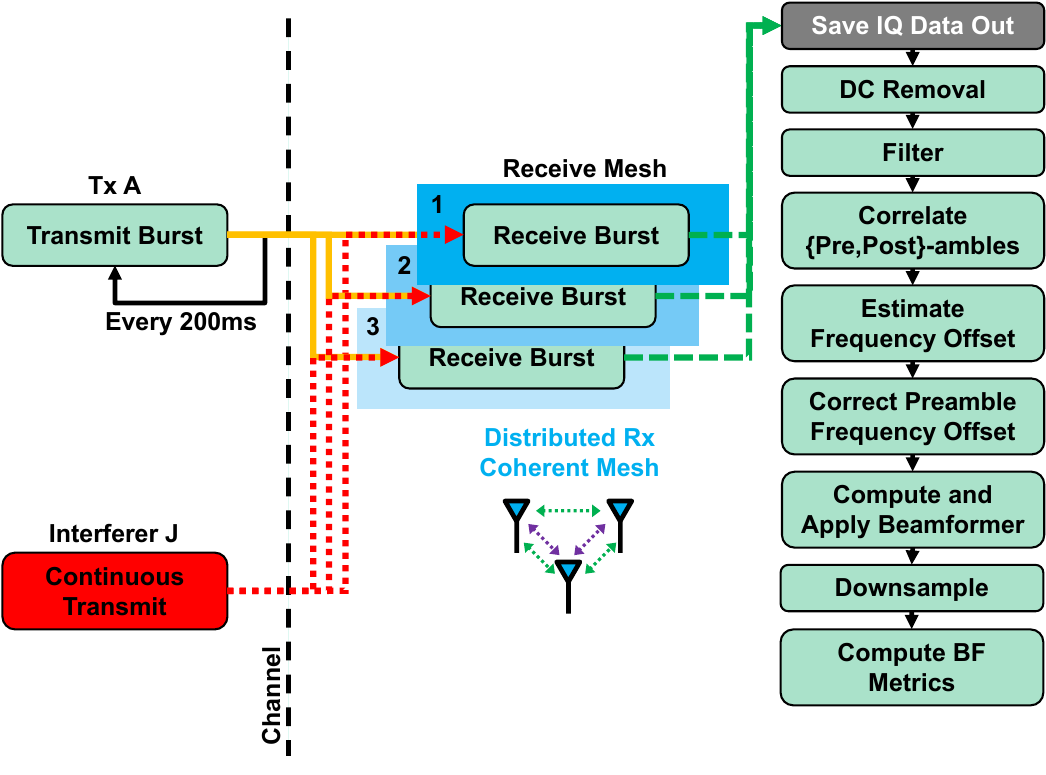}
        	\caption{Signal processing chain for the receive beamforming experiments, illustrating the flow from the transmit burst and continuous interferer through the distributed receive mesh.}
        	\label{fig:Rx_BF_Proc}
            \end{figure}

            A high-level overview of the baseband processing completed during the receive beamforming experiments is depicted in Fig. \ref{fig:Rx_BF_Proc}. Transmission bursts occur every 200 ms. 
            Upon filling the reception buffer, each node immediately passes the data to a central processing location. 
            We process receptions frame-by-frame, where we perform acquisition, preprocess the frame, estimate the beamformer, and immediately apply it to the payload section within the same frame. 
            Some preliminary steps in processing a frame include removing a DC offset through mean subtraction and performing matched filtering. 
            We acquire the signal using a normalized inner product detector with the preamble while jointly testing various time lag and CFO hypotheses. 
            Although mesh nodes are closely time-frequency aligned, a CFO between the source transmitter and each mesh node undoubtedly exists, which may arise from motion and LO frequency mismatch because the same synchronization procedure is not performed with the source. 
            This further CFO estimate coarsely corrects for a time-varying CFO frame-to-frame. We apply a common CFO correction to all received data.

            Assuming a confident detection of the signals' arrival, the last pre-processing step involves correcting any remaining CFO in the preamble at a fine resolution. 
            Training sequences are long, so to ensure an accurate channel estimate embedded in the evaluation of the beamformer, CFO in the preamble must be reduced as much as possible. 
            We obtain the CFO estimate, $\hat{f}$, using a maximum likelihood approach, which is summarized by 
            \begin{align}
                \hat{f} = \underset{f\in\mathcal{F}}{\arg \max} \quad \left\vert\sum_{t=0}^T z_n[t-\hat{\tau}] \, s^*[t] \, e^{-i2\pi f t}\right\vert^2\,,
                \label{eq:MLCFOEst}
            \end{align}
            where $\mathcal{F}$ is a discrete, finite set of hypotheses CFOs.
            With a CFO-corrected observation of the preamble, the system computes the beamformer and applies it to the current cycle's data, producing the beamformed signal. 
            The signal is then filtered and decimated to match the critical sampling rate. 
            Finally, the beamformed metrics of interest are calculated. This process repeats for all cycles.

    \subsection{Transmit Beamforming}
            \begin{figure}[!b]
            \centering
            \includegraphics[width=\linewidth]{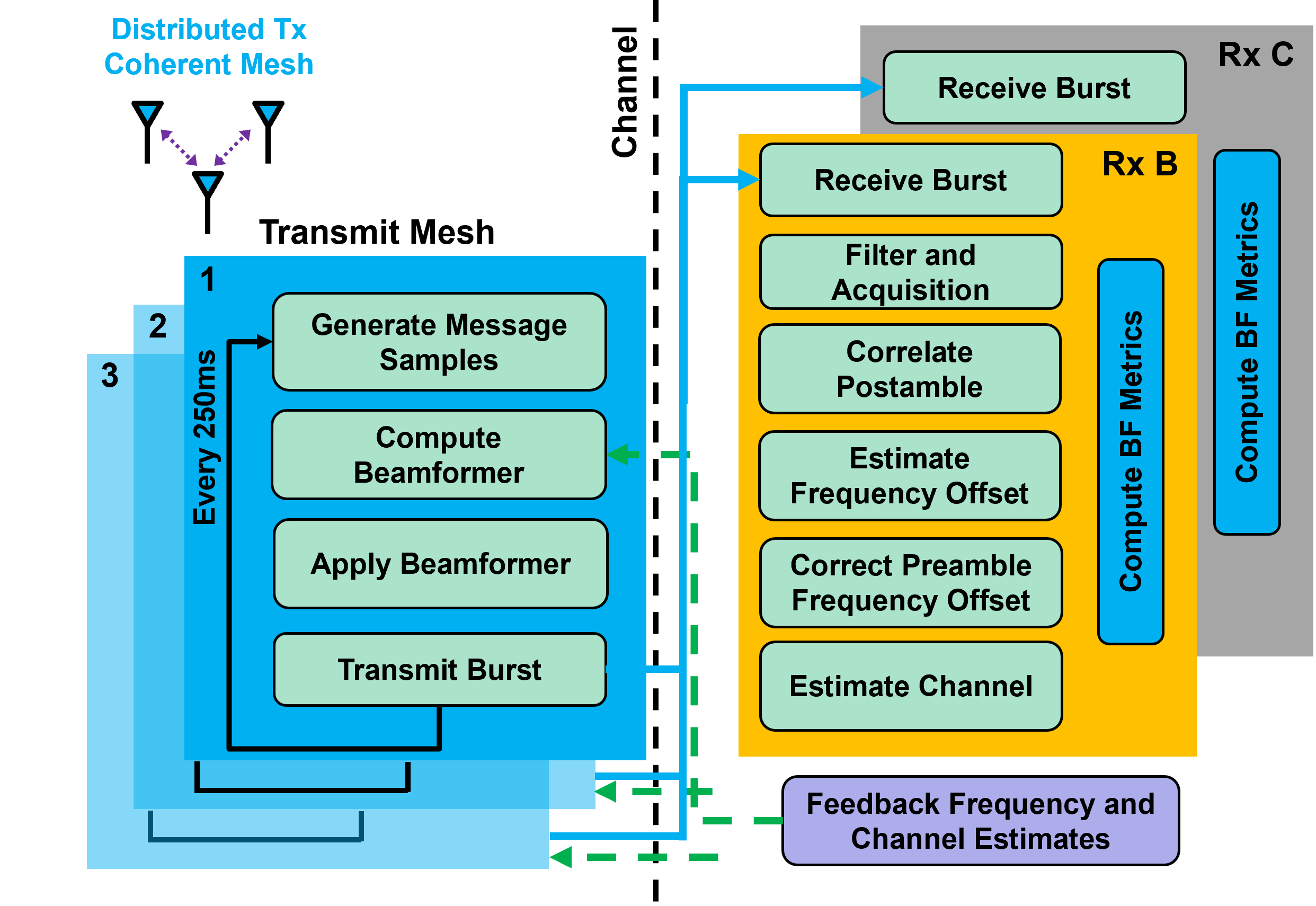}
        	\caption{Signal processing chain for the transmit beamforming experiments. 
            The transmit mesh and receivers form a feedback loop, where the latter estimates and returns channels to the former to inform beamformer construction for the subsequent transmission.
            }
        	\label{fig:Tx_BF_Proc}
            \end{figure}

            A high-level overview of the baseband processing for the transmit beamforming experiments can be found in Fig. \ref{fig:Tx_BF_Proc}. Transmission bursts occur every 250 ms. In this setup, receivers must process their receptions completely in real-time to provide the mesh with the required information to update beamformers at the burst rate. 
            Upon filling the reception buffer, the radios apply a coarse frequency correction to the data cycle-by-cycle.
            Receivers then perform joint time and frequency acquisition using a normalized inner product detector and further refine the CFO estimate through (\ref{eq:MLCFOEst}) using the postamble. Based on these CFO estimate, they apply corresponding corrections to the received preambles and obtain wideband channel estimates. 
            Finally, the receivers relay the channel estimates back to the mesh nodes. 
            We establish feedback through a side channel using a Pub-Sub messaging pattern with ZeroMQ \cite{ZeroMQ}. This messaging pattern allows each receive node to publish information through TCP while the mesh nodes subscribe to seamlessly receive the necessary information. 
            For the following cycle, the mesh elements use this channel estimate to construct a spatiotemporal matched filter beamformer and apply the beamformer before transmission. This process continues until the end of the experiment.

            \subsection{Performance Metric Calculation} \label{sec:practicalMetric}
            
                We assess beamforming performance for these experiments based on  \ac{SNR} gains, \ac{SINR} improvements, and \ac{INR} reductions. Given the known structure of the transmitted waveforms and corresponding look-through periods, we can measure all the signal powers needed to derive all \ac{SNR}, \ac{INR} and \ac{SINR} estimates. 
 
                Evaluating the $\ell^2$-norm squared of a signal segment and dividing by the number of samples yields an average signal power estimate. 
                Estimating \ac{SNR} involves evaluating $\hat{\text{SNR}} = (\hat{P}_{S+I+N} - \hat{P}_{I+N})/(\hat{P}_N)$, 
                where $\hat{P}_{S+I+N}$ is the signal-plus-interference-plus-noise power and $\hat{P}_{I+N}$ is the interference-plus-noise power, both of which can be directly estimated using relevant look-through periods. We estimate noise power $\hat{P}_N$ through noise calibration.                 Similarly, \ac{INR} can be obtained by evaluating $(\hat{P}_{I+N}-\hat{P}_{N})/\hat{P}_{N}$, and \ac{SINR} can be obtained through $(\hat{P}_{S+I+N}-\hat{P}_{I+N})/\hat{P}_{I+N}$.

                To perform noise calibration, 
                we configure each node with the same parameters as those used in the experiments, including receive gain, bandwidth, sample rate, and center frequency. First, we terminate the receive ports with impedance-matched terminators to eliminate external signal interference. Then, we conduct a series of experiments, in which a channel is sampled through the terminated port. For each experiment, we calculate the noise power as the energy per sample, and average these values across all experiments to produce a reliable estimate. Lastly, we normalize this average by dividing it by the bandwidth, resulting in an estimate per Hertz.
 
                SNR gains for \ac{SIMO} and \ac{MISO} (or receive and transmit beamforming, respectively) experiments are also computed through a ratio of the beamformed \ac{SNR}, $\text{SNR}_\text{MISO}$, to the average \ac{SISO} \ac{SNR}, $\overline{\text{SNR}}_\text{SISO}$. For example, the explicit expression for this gain in the MISO experiment is $\text{SNR Gain} = \text{SNR}_\text{MISO}/\overline{\text{SNR}}_\text{SISO}$. 
                In the transmit nulling experiments, we expect the SNR gain to be $<1$ indicating attenuation instead. 

%% file: sections/5_sec.tex
\section{Results}
We present results indicating the coherence stability offered by the entire system, then transmit beamforming and receive beamforming. Additionally, we present results of simpler transmit and receive beamforming experiments that do not require interference mitigation, demonstrating a baseline indication of success. Furthermore, we first conducted experiments with the \ac{OTS} sub-system connected through fiber for experimental ease. 
Motivated by situations in which this is impractical, such as deployment onto mobile platforms, we also reconfigured the \ac{OTS} system to exchange information through free-space
and repeated the experiments.

Drawing direct comparisons between the performances documented in this work when operating the \ac{OTS} sub-system connected via fiber optics or free-space must be avoided in this work because the state of the system and environment, namely the \ac{SISO} SINRs and INRs, differed. 
A stronger SINR yields a more accurate channel estimate, and a stronger INR yields a more accurately estimated covariance matrix. 

Meanwhile, we observe differences when comparing known theoretical values with measured quantities, such as \ac{SNR} gain. These degradations may be attributed to variations in hardware, estimation error and imbalanced reception strengths. The results presented here demonstrate that the system can achieve the expected general behavior in either configuration.

\subsection{Coherence Stability}

To observe the system's coherence stability, we perform a transmit beamforming experiment without nulling. The experiment begins by optimizing the transmission using the spatiotemporal matched filter approach in (\ref{eq:STMF}). 
The system then disables feedback from the primary receiver after 2.5 seconds to each node within the distributed mesh. Each node within the mesh then continually uses the beamforming filters computed most recently while all other system operations continue as normal. 

\begin{figure}[t]
  \centering
  \includegraphics[width=0.9\linewidth]{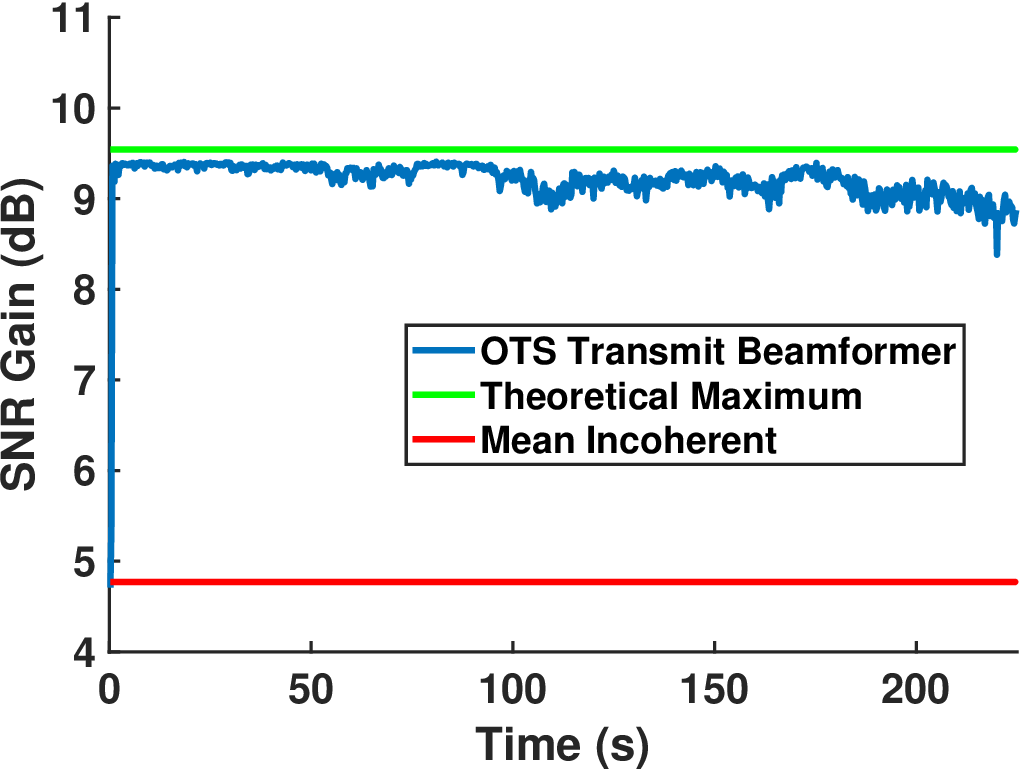}
  \caption{\ac{SNR} gain achieved during a modified transmit beamforming experiment demonstrating coherence time. 
  The system intentionally halts optimization after 2.5 s, and continually uses the most recently constructed beamformer. 
  The system initially realizes a gain approaching the theoretical bound, but degrades slowly over time as the mesh loses coherence.}
  \label{fig:CoTime}
\end{figure}

We measure the \ac{SNR} gain over time and plot the observations in Fig.~\ref{fig:CoTime}. 

The system initially closely approaches this performance bound, which is 9.5 dB for three nodes. As the elements begin to lose coherence over time, we observe a gradual degradation in the achieved gain. 
Once beamforming optimization halts, the performance starts degrading over time. This may be attributed to system channel phenomenon. 
However, the realized \ac{SNR} gain remains within 1 dB of the maximum achieved over 225 s, demonstrating the \ac{OTS} system's stability.  

\subsection{Receive Beamforming}
To test receive beamforming, we first conduct a simplified experiment, consisting of the scenario presented in Fig.~\subref{fig:IntroRx} but excluding the interferer. In this case, the \ac{OTS} uses fiber links. 
The system uses the beamformer construction in (\ref{eq:MMSEWideband}), and the covariance matrix is built from data free of the signal of interest (amounting to a covariance estimate of the noise signals in this case). 
Given a mesh size of three nodes, we expect an \ac{SNR} gain of 4.8 dB over the average SISO link. 
The time-series average \ac{SNR} gain when optimizing the beamformer is 4.1 dB, within 1 dB of the bound. 
 
Still using fiber-connected \ac{OTS} sub-sytem, we evaluate the beamforming network's performance under the exact scenario presented in Fig.~\subref{fig:IntroRx}. 
Beamforming increases the time-averaged SINR by 14.3 dB over the average SISO SINR. 
After beamforming, the system achieves a time-averaged INR of 13.5 dB below the average SISO INR. 

\begin{figure}[!t]
  \centering
\subfloat[]{\label{fig:RXBF_SINR_FS}
  \includegraphics[width=0.9\linewidth]{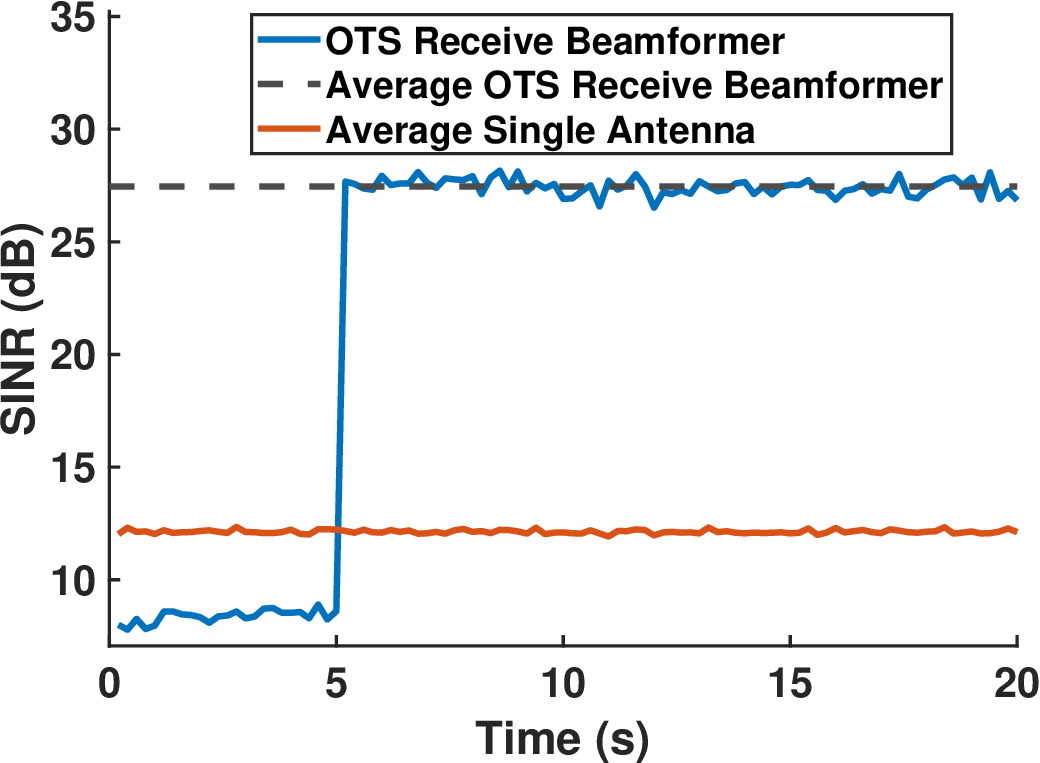}
}

\subfloat[]{\label{fig:RXBF_INR_FS}
  \includegraphics[width=0.9\linewidth]{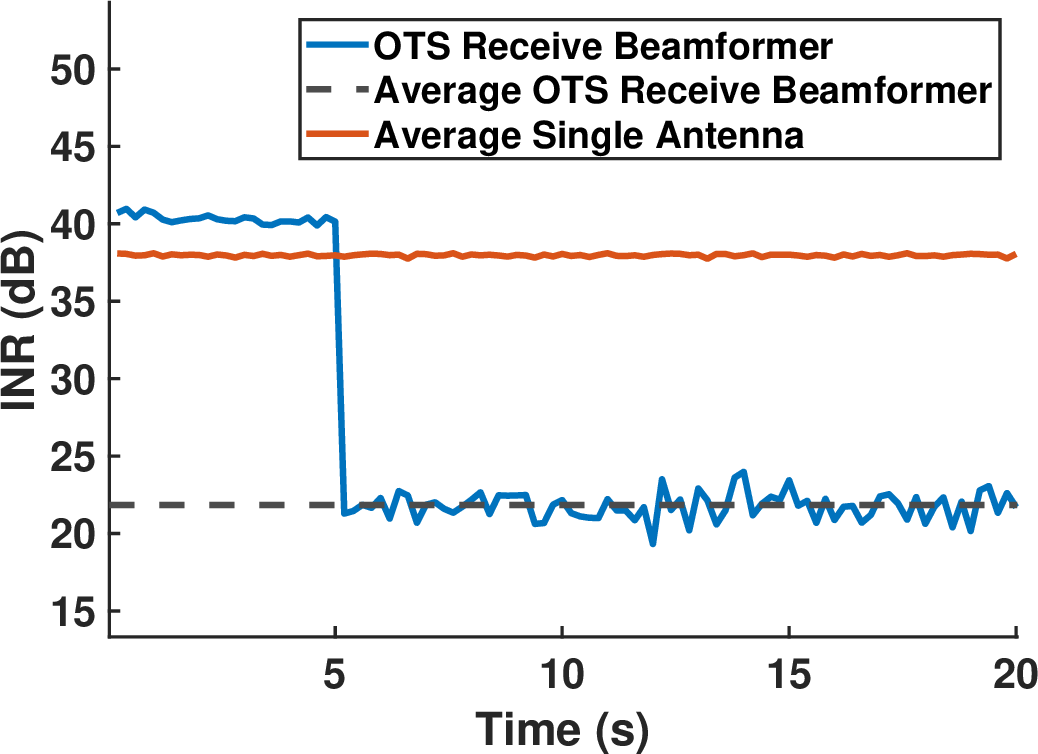}
}

  \caption{Receive Beamforming with Interference Mitigation experiment results when \ac{OTS} communicates over free space. The system (a) improves the SINR by 15.3 dB and (b) decreases the INR by 16.1 dB relative to the average SISO.}
\label{fig:RXBF_FSO}
\end{figure}

Lastly, we test the system's ability to perform receive beamforming with interference rejection, while operating the \ac{OTS} using free-space links. The SINR of the beamformed and average SISO links are plotted in Fig.~\subref{fig:RXBF_SINR_FS}. An all-1s beamformer is applied for the first 5 s of the experiment. The performance during this period is a function of the channels and the coherence state between the distributed nodes without aid from the beamformer. The time-averaged SINR of the beamformed result is 15.3 dB greater than the time-averaged SISO result. Furthermore, when examining the measured INR plotted in Fig.~\subref{fig:RXBF_INR_FS}, we observe a time-averaged reduction following beamforming of 16.1 dB. 

\subsection{Transmit Beamforming}

Similarly, to begin testing transmit beamforming, we conduct a simplified experiment, where the mesh does not optimize for the protection of a secondary receiver while the \ac{OTS} sub-system communicates over fiber. The system uses the beamformer construction in (\ref{eq:STMF}). The transmit beamforming scenario's theoretical \ac{SNR} gain is 9.5 dB for a three-node mesh. 
Once feedback is provided, the transmit beamforming network realizes an average \ac{SNR} gain of 9.3 dB. The slight difference of 0.2 dB indicates that the system successfully manipulates the signal so that phase coherence is achieved in real-time upon arrival at the receiver. 

We now test the system's ability to simultaneously generate a null toward a secondary receiver utilizing the beamformer construction in (\ref{eq::Trans_Null_MMSE_Mod}). 
The time-series average \ac{SNR} gain realized at the primary receiver (Rx B) is 7.9 dB.
Meanwhile, at secondary receiver (Rx C), an average \ac{SNR} gain of -8.3 dB is measured. This experiment demonstrates that the system can simultaneously realize an improved SNR at the intended receiver while reducing the power incident at a separate receiver.

\begin{figure}[ht]
  \centering
  \includegraphics[width=0.9\linewidth]{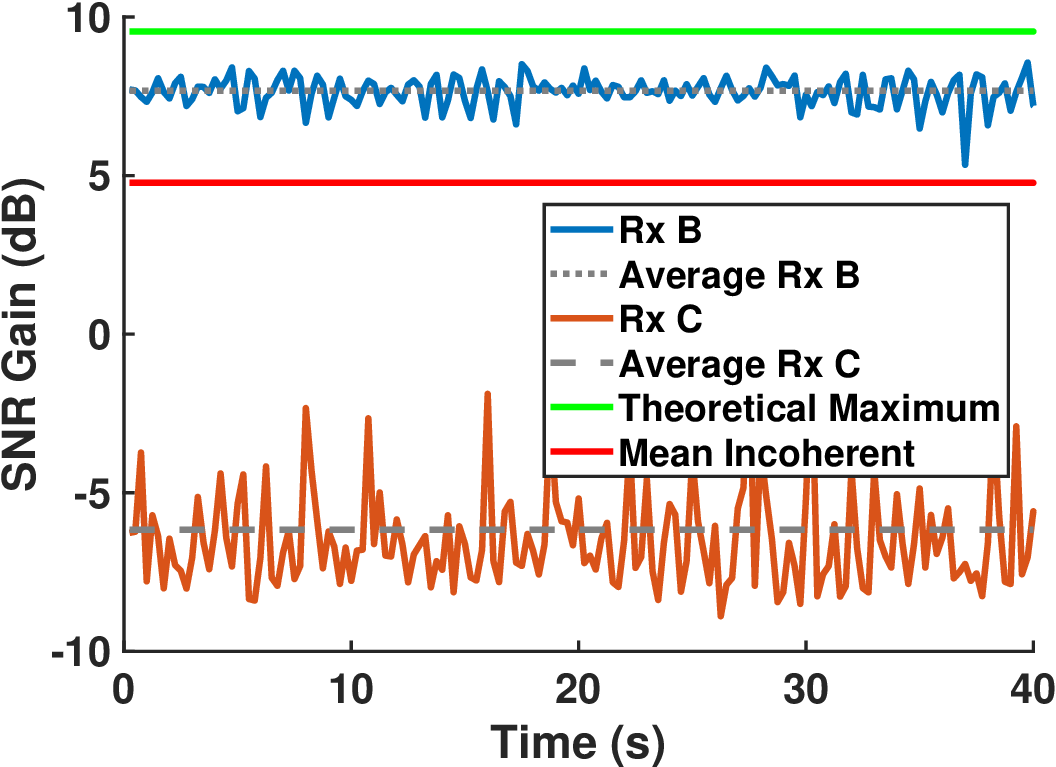}
  \caption{
  SNR Gain over average SISO SNR achieved at each receiver, Rx B and Rx C, during the transmit nulling experiment when \ac{OTS} communicates using free space. The system achieves an average SNR gain of 7.7 dB at receiver B, and -6.2 dB at receiver C, indicating both goals were accomplished.
  }
  \label{fig:TX_NULL_SNR_GAIN_FS}
\end{figure}

We conduct the transmit nulling experiment once again while instead running the \ac{OTS} with free-space connections. \ac{SNR} gains measured at both receivers are plotted in Fig.~\ref{fig:TX_NULL_SNR_GAIN_FS}. The time-series average of the \ac{SNR} gain measured at Rx B is 7.7 dB. While providing this gain, a time-average attenuation of 6.2 dB is measured at the secondary receiver.

In the transmit beamforming experiments involving nulling, we observe a degradation in the achieved \ac{SNR} gain for the intended receiver. This observation may partly be attributed to insufficient \ac{DoF} or channel diversity. Purely spatial beamformer constructions were used only in transmit beamforming experiments, requiring nulling. Thus, only $N$ \ac{DoF} were available. 
Not all channels may be independent, reducing the number of useful \ac{DoF}, thus degrading the overall potential efficacy of the beamformer. 

%% file: sections/6_conc.tex
\section{Conclusion}

    In this study, we constructed a distributed coherent beamforming system that operates at a carrier frequency of 60 GHz. We successfully demonstrated its ability to significantly improve the receive power of communications signals and simultaneously suppress interference under both receive and transmit modalities. 
    To enable this capability, we achieved precise phase synchronization across multiple elements within the distributed network by integrating an optical time synchronization network into the system, implementing an RF timestamp synchronization technique, applying fine frequency correction, and constructing adaptive beamformers. 
    
    The system achieves expected beamforming behavior when the \ac{OTS} system exchanges information over free-space and fiber. The results demonstrate the feasibility of achieving distributed coherent beamforming at 60 GHz and the potential of deploying this system for advanced wireless communication systems. 

    These promising results provide the foundation for tackling new challenges. Implementing an optical coherence-based distributed system requires careful consideration of cost and scalability. The optical system's pointing and tracking requirements present potential limitations in large-scale deployment, though this may be solved by the proliferation of free-space optical communication. Furthermore, RF-based coherence techniques should be explored at mmWave frequencies as an alternative.
    Given the accuracy of the \ac{OTS} system, higher frequency applications, such as terahertz communication, are one viable avenue of further research. 
    We are also exploring calibration techniques as an alternative to waveform exploitation to enable open-loop operations, such as beam-steering. Additional high-level applications, such as multi-static distributed radar and distributed direction finding, remain open research interests.